\documentclass[prb,twocolumn,superscriptaddress,showpacs,amsmath,amssymb]{revtex4}
\usepackage{amsfonts}
\usepackage{bbm}
\usepackage{graphicx}
\usepackage{dcolumn}
\usepackage{bm}
\usepackage{color}

\begin{document}
\title{Spin-valley dependent double Andreev reflections in the proximitized graphene/superconductor junction}

\author{Lu Gao}
\affiliation{School of Science, Qingdao University of Technology, Qingdao, Shandong 266520, China}
\author{Qiang Cheng}
\email[]{chengqiang07@mails.ucas.ac.cn}
\affiliation{School of Science, Qingdao University of Technology, Qingdao, Shandong 266520, China}
\affiliation{International Center for Quantum Materials, School of Physics, Peking University, Beijing 100871, China}

\author{Qing-Feng Sun}
\email[]{sunqf@pku.edu.cn}
\affiliation{International Center for Quantum Materials, School of Physics, Peking University, Beijing 100871, China}
\affiliation{Collaborative Innovation Center of Quantum Matter, Beijing 100871, China}
\affiliation{Hefei National Laboratory, Hefei 230088, China}

\begin{abstract}
We study the Andreev reflections and the quantum transport in the proximitized graphene/superconductor junction. The proximitized graphene possesses the pseudospin staggered potential and the intrinsic spin-orbit coupling induced by substrate, which are responsible for the spin-valley dependent double Andreev reflections and the anomalous transport properties in the junction.
The pure specular Andreev reflection can happen in the superconducting gap
for the $K\uparrow$ and $K'\downarrow$ electrons
while the pure retro-Andreev reflection happens for the $K\downarrow$ and $K'\uparrow$ electrons.
The coexisting two types of Andreev reflections related to the fixed spin-valley indices strongly depend on the chemical potential of the proximitized graphene. The condition of the emergence of the specific type of Andreev reflection for the electrons with the fixed spin-valley index is clarified.
The spin-valley dependent Andreev reflections bring about the peculiar conductance spectra of the junction, which can help determine the values of the pseudospin staggered potential and the intrinsic spin-orbit coupling induced in graphene. Hence, our research results not only provide an experimental method to detect the induced potential and coupling in graphene
but also establish the foundation of the superconductor electronics
based on the spin-valley indices.
\end{abstract}
\maketitle
\section{\label{sec1}Introduction}
The combination of the Andreev reflection and the degrees of freedom of
electrons can help establish novel electronics based on superconductor.
The Andreev reflection is a basic physical process in the condensed matter physics, which dominates the low-energy quantum transport in
normal conductor/superconductor heterostructures.
For the conventional metal/superconductor junction,
the Andreev reflected hole moves back along the trajectory of
the incident electron from the metal. This type of Andreev reflection
is called the retro-Andreev reflection (RAR)\cite{Andreev},
which has proved to be an effective tool for the detection of the superconducting pairing wave functions\cite{Tanaka1,Tanaka2,Takabatake,Anwar}.
However, the spin of electrons in the conventional metal is degenerate,
which limits the further studies and applications of RAR.
Different from the conventional metal, ferromagnet possesses the spin-polarized electronic structure. The spin degree of freedom
can be fully utilized in the RAR process in the ferromagnet/superconductor junctions\cite{Zutic}. The resulting spin-polarized transport\cite{Zheng,Hirai}, the long-range triplet proximity effect\cite{Buzdin1} and the various Josephson ground states\cite{Buzdin2,Goldobin,QCheng1} in the junctions together constitute the fundamentals of the superconducting spintronics\cite{Linder1}.

As the two-dimensional Dirac fermion system, graphene can host another
type of Andreev reflection, i.e., the specular Andreev reflection (SAR)\cite{Beenakker1}, in which the hole will be specularly reflected. The conversion from RAR to SAR in graphene can be tuned by changing the chemical potential with a gate voltage. The two types of Andreev reflections, RAR and SAR, and their resultant physical phenomena in the graphene/supercondcutor junction have attracted many researches\cite{Beenakker2}.
The anomalous oscillatory behavior of conductance, the pure crossed Andreev reflection and the Josephson effect based on the anisotropic pairing have been clarified\cite{Linder2,Linder3,Cayssol}. Several mechanisms have also been proposed to detect or distinguish the two Andreev reflections, which include the suppression of RAR by the ferromagnetic exchange interaction\cite{Zhang}, the period feature of the Aharonov-Bohm conductance oscillations\cite{Schelter} and the quantum interference mechanism in a four-terminal graphene-superconductor system\cite{Cheng1,Xing,Cheng2}.

However, there are still two questions to be solved in the studies on the Andreev reflection in graphene. First, both spin and valley are degenerate in graphene due to the negligible spin-orbit coupling. The Andreev reflection, including its type and its magnitude, is independent on the spin and valley indices of electrons. One can not obtain the spin-valley dependent Andreev reflections in the isolated graphene/superconductor junctions. Second, for the given chemical potential and the given energy of incident electrons, RAR and SAR will not happen simultaneously and only one of them can happen. The two questions limit the further investigation on the Andreev reflections in graphene and the establishment of the spin-valley electronics based on superconductor.
In addition, Majidi and Zareyan have studied the Andreev reflection
in the graphene superconductor/pseudoferromagnet junction,
and they demonstrate an enhanced Andreev reflection caused by the gap from the pseudospin staggered potential\cite{Majidi}.
However, the spin and the valley of electrons there are still degenerate
and only one of RAR and SAR can happen
since the intrinsic spin-orbit coupling is not considered there.
Recently, theoretical and experimental investigations demonstrate that the pseudospin staggered potential and the intrinsic spin-orbit coupling can be induced by the proximity effect of substrates such as transition metal dichalcogenides\cite{Zihlmann,Zollner1,Zollner2,Khatibi,Wakamura,Frank,Gmitra,Wang,aref1}. The two induced interactions of $meV$ order of magnitude break the quadruple degeneracy of spin and valley and favor the researches on the low-energy transport. The transport properties, the spin-valley resolved curvature and energy spectra in the proximitized graphene with the two induced interactions have been investigated\cite{Zubair,Cysne,QCheng2}.

In this paper, we propose the junction composed of the proximitized graphene and superconductor to study the combination of the spin-valley degree of freedom and the Andreev reflection. We find in the junction the Andreev reflections are spin-valley dependent. SAR occurs at the interface when the $K\uparrow$ or $K'\downarrow$ electrons are injected from graphene while RAR occurs when the $K\downarrow$ or $K'\uparrow$ electrons are injected.
The spin-valley resolved double Andreev reflections coexist in the proximitized graphene, which happen simultaneously under a fixed chemical potential for a given energy of the incident electrons. These peculiar characters are distinct from the isolated graphene/superconductor junction\cite{Beenakker1} and bring about the anomalous conductance spectra.
The two aforementioned questions encountered in the isolated graphene/supercondcutor junctions to further study the Andreev reflectoin can be solved. Both the spin-valley dependent Andreev reflections and the coexisting SAR and RAR can be realized in the same proximitized graphene/superconductor junction at once.
Furthermore, from the conductance spectra, one can determine the magnitude of the induced pseudospin staggered potential and the intrinsic spin-orbit coupling in graphene. We also investigate the influences of the chemical potential on the double Andreev reflections and the conductance. The conditions for the occurrences of SAR and RAR for the definite spin-valley indices are presented. The clarification of the spin-valley dependent Andreev reflections and the transport properties in the proximitized graphene/superconductor junction establishes the foundation of the spin-valley dependent superconductor electronics.

The organization of this paper is as follows. In Sec.\uppercase\expandafter{\romannumeral 2}, we give the model for the proximitized graphene/superconductor junction and the formalism for the reflection probabilities and conductance. In Sec.\uppercase\expandafter{\romannumeral 3}A and B,
the numerical results for the probabilities and conductance are presented and discussed in two cases with and without the staggered potential.
The dependence of the conductance on the chemical potential
and the bias voltage of the junction are studied in
Sec.\uppercase\expandafter{\romannumeral 3}C.
Sec.\uppercase\expandafter{\romannumeral 4} concludes
this paper. The detailed formula derivation process for the reflection
probabilities is put in APPENDIX.
\section{\label{sec2}Model and Formulation}
\begin{figure}[!htb]
\centerline{\includegraphics[width=0.9\columnwidth]{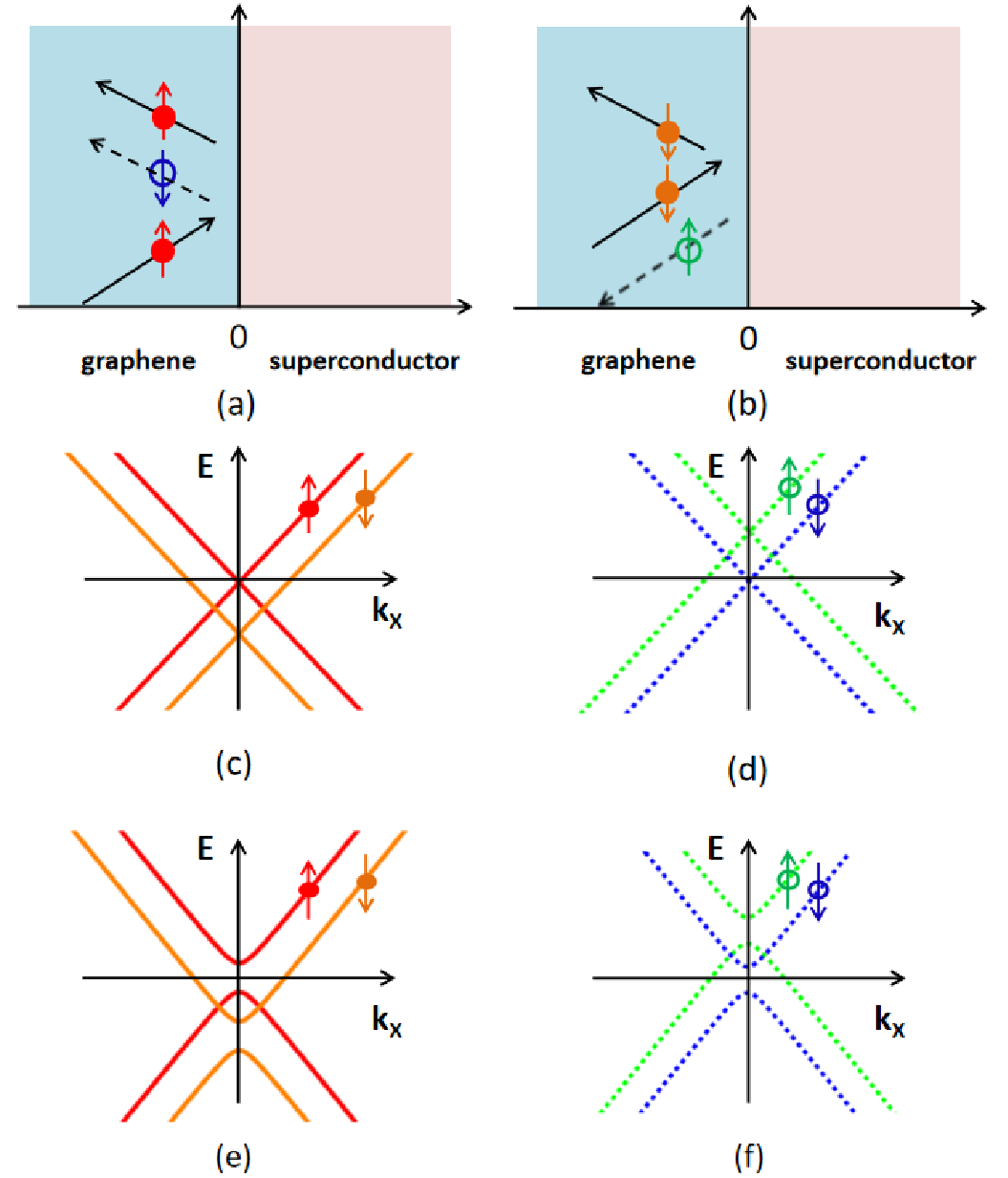}}
\caption{The scattering processes and the dispersions in the proximitized graphene. (a) SAR and the normal reflection for the incident $K\uparrow$ electrons. (b) RAR and the normal reflection for the incident $K\downarrow$ electrons. (c) The dispersions $E_{eK\uparrow}^{\pm}$ (red lines) and $E_{eK\downarrow}^{\pm}$ (orange lines) for electrons in the $K$ valley with the pseudospin staggered potential $\Delta=0$. (d) The dispersions for holes associated with the electrons in (c). (e) The dispersions $E_{eK\uparrow}^{\pm}$ (red lines) and $E_{eK\downarrow}^{\pm}$ (orange lines) for electrons in the $K$ valley with $\Delta\ne0$. (f) The dispersions for holes associated with the electrons in (e). Solid (hollow) circles denote electrons (holes). Short arrows denote spin of electrons and holes. The solid (dashed) long arrows in (a) and (b)
denote the motion direction of electrons (holes).
The scattering processes in (a) and (b) can happen at the same incident energy. Other parameters are taken as $\lambda=\mu_{g}\ge\Delta_{0}$. }
\label{fig1}
\end{figure}

The proximitized graphene/superconductor junction we consider is schematically shown in Figs.{\ref{fig1}}(a) and (b). The interface of the junction is located at $x=0$ and is parallel to the $y$ axis. The electric transport is along the $x$ axis. For the proximitized graphene, the pseudospin staggered potential $\Delta$ and the intrinsic spin-orbit coupling $\lambda$ are induced from the substrate and the Hamiltonian in the pseudospin space can be written as\cite{Gmitra,Wang,Zubair,Cysne,QCheng2}
\begin{equation}
   h_{g}=\begin{pmatrix}
     \Delta +\xi s_{z}\lambda -\mu _{g} & \hbar v_{F}(\xi k_{x}-ik_{y}  )  \\
      \hbar v_{F}(\xi k_{x}+ik_{y}) & -\Delta +\xi s_{z}\lambda -\mu _{g}
      \end{pmatrix}.\label{gH}
\end{equation}
acting on the two-dimensional electron spinor $(u_{A\xi s_{z}},u_{B\xi s_{z}})^{T}$\cite{Beenakker1}. Here $\xi=+1(-1)$ for the $K(K')$ valley, $s_{z}=+1(-1)$ for the up(down) spin of electrons, $v_{F}$ is the Fermi velocity, $\mu_{g}$ is the chemical potential, ${\bf{k}}=(k_{x},k_{y})$ is the wave vector in the $x$-$y$ plane and the subscript $A(B)$ denotes the $A(B)$ sublattice.

From Eq.(\ref{gH}), we can write the Bogoliubov-de Gennes (BdG) Hamiltonian for graphene as
\begin{equation}
H_{G}=\left(\begin{array}{cc}
h_{g}&0\\
0&-h_{g}\end{array}\right),\label{HGbdg}
\end{equation}
which satisfy the BdG equation $H_{G}\psi=E\psi$ with $k_{x}=-i\frac{\partial}{\partial x}$. The eigenvalues of the equation can be solved as
\begin{align}
E_{e\xi s_{z}}^{\pm}&=\pm\sqrt{\hbar^{2}v_{F}^{2}(k_{x}^{2}+k_{y}^{2})+\Delta^{2}}+\xi s_{z}\lambda-\mu_{g},\label{evele}\\
E_{h\xi s_{z}}^{\pm}&=\pm\sqrt{\hbar^{2}v_{F}^{2}(k_{x}^{2}+k_{y}^{2})+\Delta^{2}}-\xi s_{z}\lambda+\mu_{g},\label{evhole}
\end{align}
for electrons and holes, respectively.
Here, $E_{e\xi s_{z}}^{+}$ ($E_{h\xi s_{z}}^{-}$)
and $E_{e\xi s_{z}}^{-}$ ($E_{h\xi s_{z}}^{+}$)
are the conduction band and the valence band for electrons (holes), respectively.

For superconductor, i.e., the superconducting region with $x>0$, we consider the graphene without the stagger potential $\Delta$ and the spin-orbit coupling $\lambda$ and its superconductivity is induced by a bulk superconductor covered on it. In this region, the BdG Hamiltonian is given by\cite{Beenakker1,Asano}
\begin{equation}
H_{S}=\left(\begin{array}{cc}
h_{s}& s_{z}\Delta_{0}{\bf 1}_{2\times2}\\
 s_{z}\Delta_{0}{\bf 1}_{2\times2}&-h_{s}\end{array}\right),\label{SBdG1}
\end{equation}
with
\begin{equation}
h_{s}=\left(\begin{array}{cc}
-\mu_{s}&\hbar v_{F}(\xi k_{x}-i k_{y})\\
\hbar v_{F}(\xi k_{x}+i k_{y})&-\mu_{s}\end{array}\right),\label{SBdG}
\end{equation}
and the identity matrix ${\bf 1}_{2\times2}$. Here $\mu_{s}$ is the chemical potential in superconductor and $\Delta_{0}$ is the magnitude of the superconducting gap. In Eq.(\ref{SBdG1}), the superconducting paring potential couples the $K\uparrow(K\downarrow)$ electrons with the $K'\downarrow(K'\uparrow)$ holes or the $K'\uparrow(K'\downarrow)$ electrons with the $K\downarrow(K\uparrow)$ holes.
Both $H_{G}$ in Eq.(\ref{HGbdg}) and $H_{S}$ in Eq.(\ref{SBdG1}) act on the four-dimensional spinor $(u_{A\xi s_{z}},u_{B\xi s_{z}},v_{A\bar{\xi}\bar{s_{z}}},v_{B\bar{\xi}\bar{s_{z}}})^T$ with the two-dimensional hole spinor $(v_{A\bar{\xi}\bar{s_{z}}},v_{B\bar{\xi}\bar{s_{z}}})^T$\cite{Beenakker1}. Here, $\bar{\xi}=-\xi$ and $\bar{s_{z}}=-s_{z}$.
The BdG equation for superconductor is given by $H_{S}\psi=E\psi$ with $k_{x}=-i\frac{\partial}{\partial x}$.
From the BdG equation, the eigenvalues for superconductor can be solved as $E^{\pm}=\pm\sqrt{(\hbar v_{F}k-\mu_{s})^2+\Delta_{0}^2}$.

Since the $K\uparrow(K\downarrow)$ and the $K'\downarrow(K'\uparrow)$ electrons obey the time-reversal symmetry, we will next only consider the electrons in the $K$ valley. Here we take the situation of $\lambda=\mu_{g}$ as an example to present the derivation of the expressions of reflection probabilities and conductance (see APPENDIX for other situations). In this case, both the $K\uparrow$ and the $K\downarrow$ electrons participating in transport come from their conduction bands.
For the pseudospin staggered potential $\Delta=0$, both the $K\uparrow$ and $K\downarrow$ electrons and their associated holes in the proximitized graphene have the linear dispersions as schematically shown in Figs.{\ref{fig1}}(c) and (d). For $\Delta\ne0$, gaps will open in the dispersions as schematically shown in Figs.{\ref{fig1}}(e) and (f).

For an incident $K\uparrow$ electron, SAR happens in the incident energy range $E>0$ and the Andreev reflected hole is in its valence band.
For an incident $K\downarrow$ electron, RAR happens in the range $0<E<\mu_{g}+\lambda$
and the Andreev reflected hole is in its conduction band,
while SAR happens in the range $E>\mu_{g}+\lambda$ and
the Andreev reflected hole is in its valence band.
Taking $\mu_{g}+\lambda\ge\Delta_{0}$ into account,
pure SAR for the $K\uparrow$ electrons and pure RAR for the $K\downarrow$ electrons are expected in the superconducting gap with $E<\Delta_{0}$ which is the dominate energy range for Andreev reflections. They are the spin-valley dependent double Andreev reflections in the proximitized graphene/superconductor junction.
It needs to be emphasized that the two types of Andreev reflections
(RAR and SAR) occur at the same incident energy $E$.
The scattering processes are schematically shown separately in Figs.{\ref{fig1}}(a) and (b).

For the injection of a $K\uparrow$ electron from the conduction band, the wave function in graphene with $x<0$ can be solved from the BdG equation as
\begin{eqnarray}
\begin{split}
\psi_{G}^{K\uparrow}&=\left(\begin{array}{c}
\chi_{1}\eta_{e\uparrow+} \\
1 \\
0\\
0\end{array}\right)e^{i k_{x}^{e\uparrow+}x}
+r_{1}\left(\begin{array}{c}
-\chi_{1}\eta_{e\uparrow+}^{*}\\
1\\
0\\
0\end{array}\right)e^{-i k_{x}^{e\uparrow+}x}\\
&+
r_{a1}\left(\begin{array}{c}
0\\
0\\
\chi_{2}\eta_{h\downarrow+}\\
1\end{array}\right)e^{-i k_{x}^{h\downarrow+}x},\label{wfku}
\end{split}
\end{eqnarray}
with $\chi_{1}=(E-\lambda+\Delta+\mu_{g})/\sqrt{(E-\lambda+\mu_{g})^2-\Delta^2}$, $\chi_{2}=(E+\lambda-\Delta-\mu_{g})/\sqrt{(E+\lambda-\mu_{g})^2-\Delta^2}$, $\eta_{e\uparrow+}=(k_{x}^{e\uparrow+}-i k_{y})/k_{e\uparrow+}$,
$\eta_{h\downarrow+}=(k_{x}^{h\downarrow+}+i k_{y})/k_{h\downarrow+}$, $k_{x}^{e\uparrow+(h\downarrow+)}=\sqrt{[E-(+)\lambda+(-)\mu_{g}]^2-\Delta^2-\hbar^2v_{F}^2k_{y}^2}/{\hbar v_{F}}$ and $k_{e\uparrow+(h\downarrow+)}=\sqrt{\left({k_{x}^{e\uparrow+(h\downarrow+)}}\right)^2+k_{y}^2}$. Here, $k_{y}=\left[\sqrt{(E-\lambda+\mu_{g})^2-\Delta^2}/{\hbar v_{F}}\right]\sin{\theta}$ with the incident angle $\theta$ of electrons.
$r_{a1}$ and $r_1$ in Eq.(\ref{wfku}) represent the SAR coefficient
and the normal reflection coefficient, respectively.

For the injection of a $K\downarrow$ electron from the conduction band, the wave function in graphene with $x<0$ can be solved as
\begin{eqnarray}
\begin{split}
\psi_{G}^{K\downarrow}&=\left(\begin{array}{c}
\chi_{3}\eta_{e\downarrow+}\\
1\\
0\\
0\end{array}\right)e^{i k_{x}^{e\downarrow+}x}+r_{2}\left(\begin{array}{c}
-\chi_{3}\eta_{e\downarrow+}^{*}\\
1\\
0\\
0\end{array}\right)e^{-i k_{x}^{e\downarrow+}x}\\
&+r_{a2}\left(\begin{array}{c}
0\\
0\\
-\chi_{4}\eta_{h\uparrow-}\\
1\end{array}\right)e^{i k_{x}^{h\uparrow-}x},\label{wfkd1}
\end{split}
\end{eqnarray}
for $0< E<\mu_{g}+\lambda$ and
\begin{eqnarray}
\begin{split}
\psi_{G}^{K\downarrow}&=\left(\begin{array}{c}
\chi_{3}\eta_{e\downarrow+}\\
1\\
0\\
0\end{array}\right)e^{i k_{x}^{e\downarrow+}x}+r_{2}\left(\begin{array}{c}
-\chi_{3}\eta_{e\downarrow+}^{*}\\
1\\
0\\
0\end{array}\right)e^{-i k_{x}^{e\downarrow+}x}\\
&+r_{a2}\left(\begin{array}{c}
0\\
0\\
\chi_{4}\eta_{h\uparrow+}\\
1\end{array}\right)e^{-i k_{x}^{h\uparrow+}x},\label{wfkd2}
\end{split}
\end{eqnarray}
for $E>\mu_{g}+\lambda$. Here, $r_{a2}$ in Eq.(\ref{wfkd1}) represents the RAR coefficient while $r_{a2}$ in Eq.(\ref{wfkd2}) represents the SAR coefficient.
The expressions of the symbols in wave functions are $\chi_{3}=(E+\lambda+\Delta+\mu_{g})/\sqrt{(E+\lambda+\mu_{g})^2-\Delta^2}$,
$\chi_{4}=(E-\lambda-\Delta-\mu_{g})/\sqrt{(E-\lambda-\mu_{g})^2-\Delta^2}$,
$\eta_{e\downarrow+}=(k_{x}^{e\downarrow+}-i k_{y})/{k_{e\downarrow+}}$, $\eta_{h\uparrow-}=(k_{x}^{h\uparrow-}-i k_{y})/{k_{h\uparrow-}}$, $\eta_{h\uparrow+}=(k_{x}^{h\uparrow+}+i k_{y})/k_{h\uparrow+}$, $k_{x}^{e\downarrow+(h\uparrow-)}=\sqrt{[E+(-)\lambda+(-)\mu_{g}]^2-\Delta^2-\hbar^2v_{F}^2k_{y}^2}/\hbar v_{F}$ and $k_{e\downarrow+(h\uparrow-)}=\sqrt{\left({k_{x}^{e\downarrow+(h\uparrow-)}}\right)^2+k_{y}^2}$ with $k_{y}=\left[\sqrt{[E+\lambda+\mu_{g}]^2-\Delta^2}/\hbar v_{F}\right]\sin{\theta}$. The wave vectors $k_{x}^{h\uparrow+}$ and $k_{h\uparrow+}$ are the same with $k_{x}^{h\uparrow-}$ and $k_{h\uparrow-}$, respectively.

In superconductor, the wave function with $x>0$ for the injection of a $K\uparrow$ electron can be given by
\begin{eqnarray}
\psi_{S}^{K\uparrow}=t_{1}\left(\begin{array}{c}
u\eta\\
u\\
v\eta\\
v\end{array}\right)e^{i k_{x}x}
+t_{2}\left(\begin{array}{c}
-v\eta^{*}\\
v\\
-u\eta^{*}\\
u\end{array}\right)e^{-i k_{x}x},\label{wfs}
\end{eqnarray}
and that for a $K\downarrow$ electron is given by
\begin{eqnarray}
\psi_{S}^{K\downarrow}=t_{1}\left(\begin{array}{c}
-u\eta\\
-u\\
v\eta\\
v\end{array}\right)e^{i k_{x}x}
+t_{2}\left(\begin{array}{c}
v\eta^{*}\\
-v\\
-u\eta^{*}\\
u\end{array}\right)e^{-i k_{x}x},\label{wfs}
\end{eqnarray}
with $u=\sqrt{(E+\Omega)/2E}$, $v=\sqrt{(E-\Omega)/2E}$, $\Omega=\sqrt{E^2-\Delta_{0}^2}$, $\eta=(k_{x}-ik_{y})/k$, $k_{x}=\sqrt{\mu_{s}^2-\hbar^2 v_{F}^2k_{y}^2}/\hbar v_{F}$ and $k=\sqrt{k_{x}^2+k_{y}^2}$.

Using the following boundary conditions,
\begin{eqnarray}
 \psi_{G}^{K\uparrow}(x=0^{-})=\psi_{S}^{K\uparrow}(x=0^{+}),\\
 \psi_{G}^{K\downarrow}(x=0^{-})=\psi_{S}^{K\downarrow}(x=0^{+}),
\end{eqnarray}
the normal reflection coefficients $r_{1}$, $r_{2}$ and the Andreev reflection coefficients $r_{a1}$, $r_{a2}$ can be derived.
Then, the probabilities for the normal reflections are given by $R_{1}=\vert r_{1}\vert^2$ and $R_{2}=\vert r_{2}\vert^2$.
The probability of SAR for the incident $K\uparrow$ electrons is given by\cite{Hou,QCheng3}
\begin{eqnarray}
R_{a1}=\frac{\chi_{2}k_{x}^{h\downarrow+}k_{e\uparrow+}}{\chi_{1}k_{x}^{e\uparrow+}k_{h\downarrow+}}\vert r_{a1}\vert^2.\label{Ra1}
\end{eqnarray}
The probability of RAR in the energy range $E<\mu_{g}+\lambda$ for the $K\downarrow$ electrons is given by
\begin{eqnarray}
R_{a2}=-\frac{\chi_{4}k_{x}^{h\uparrow-}k_{e\downarrow+}}{\chi_{3}k_{x}^{e\downarrow+}k_{h\uparrow-}}\vert r_{a2}\vert^2,\label{Ra21}
\end{eqnarray}
and that of SAR in the range $E>\mu_{g}+\lambda$ for the $K\downarrow$ electrons is given by
\begin{eqnarray}
R_{a2}=\frac{\chi_{4}k_{x}^{h\uparrow+}k_{e\downarrow+}}{\chi_{3}k_{x}^{e\downarrow+}k_{h\uparrow+}}\vert r_{a2}\vert^2.\label{Ra22}
\end{eqnarray}
The defined probabilities satisfy the conservation conditions $R_{1}+R_{a1}=1$ and $R_{2}+R_{a2}=1$ when the incident energy $E<\Delta_{0}$.

According to the Blonder-Tinkham-Klapwijk theory\cite{BTK},
the normalized conductance for the bias voltage $V$ at the zero temperature can be correspondingly expressed as
\begin{align}
\sigma_{K\uparrow}=&\frac{e^{2}}{h}\frac{N_{\uparrow}(eV)}{\sigma_{0}}\int_{-\frac{\pi}{2}}^{\frac{\pi}{2}}(1+R_{a1}-R_{1})\cos{\theta}d\theta,\label{cond1}\\
\sigma_{K\downarrow}=&\frac{e^{2}}{h}\frac{N_{\downarrow}(eV)}{\sigma_{0}}\int_{-\frac{\pi}{2}}^{\frac{\pi}{2}}(1+R_{a2}-R_{2})\cos{\theta}d\theta,\label{cond2}
\end{align}
with the number of the transverse modes $N_{\uparrow(\downarrow)}(E)=[W\sqrt{(E-(+)\lambda+\mu_{g})^{2}-\Delta^{2}}]/{\pi\hbar v_{F}}$ for the width $W$ of graphene. The conductance $\sigma_{0}=\sigma_{K\uparrow0}+\sigma_{K\downarrow0}$ with $\sigma_{K\uparrow0}$ and $\sigma_{K\downarrow0}$ denoting conductances for the $K\uparrow$ electrons and the $K\downarrow$ electrons when the superconductor is in the normal state. Hence, the normalized total conductance can be written as
\begin{eqnarray}
\sigma=\sigma_{K\uparrow}+\sigma_{K\downarrow}.\label{cond}
\end{eqnarray}
From Eqs.(\ref{cond1})-(\ref{cond}), one can find that conductances are determined by the modes and reflection probabilities concurrently, which are functions of the bias $V$ for the given spin-orbit coupling strength $\lambda$, staggered potential $\Delta$ and chemical potential $\mu_{g}$. When writing Eq.(\ref{cond}), the degeneracy of the $K\uparrow(K\downarrow)$ electrons and the $K'\downarrow(K'\uparrow)$ electrons has been involved.
\section{\label{sec3}Results and discussions}
In our calculations, we have taken the magnitude of the superconducting gap $\Delta_{0}$ as the unit of energy. The wave vector $k_{0}=\Delta_{0}/\hbar v_{F}$ is defined to normalize the wave vectors in wave functions. The chemical potential $\mu_{s}$ in superconductor is taken as $\mu_{s}=10\Delta_{0}$ and the intrinsic spin-orbit coupling $\lambda$ is taken as $\lambda=0.5\Delta_{0}$ in Secs.\uppercase\expandafter{\romannumeral 3}A and \uppercase\expandafter{\romannumeral 3}B or $\lambda=0.2\Delta_{0}$ in Sec.\uppercase\expandafter{\romannumeral 3}C. Taking the $meV$ order of magnitude for $\Delta_{0}$ into account, the taken values for $\lambda$ here are realistic\cite{Gmitra}. For the Andreev reflections and their probabilities, we will focus on the energy in the superconducting gap with $0<E/\Delta_{0}<1$ since it is the dominate range of their occurrence.

\subsection{$\Delta=0$}
In this subsection, we consider the proximitized graphene without
the pseudospin staggered potential, i.e., $\Delta=0$.
In this situation, the dispersions of electrons and holes in graphene
are linear. According to the eigenvalues of $H_{G}$ presented in Eq.(\ref{evele}), the effective chemical potential $\mu_{eff}$ for the $K\uparrow$ electrons is $(\mu_{g}-\lambda)$ and that for the $K\downarrow$ electrons is $(\mu_{g}+\lambda)$, which will determine the type of the Andreev reflection for electrons with the specific spin-valley index.
For the isolated graphene with $\lambda=\Delta=0$,
the electrons have the quadruple degeneracy of spin and valley.
The electrons with the different spin and valley indices possess
the same dispersions and the same chemical potential.
Then only one of RAR and SAR occurs for a fixed incident energy.
The conversion from RAR to SAR is achieved by adjusting the chemical
potential or the energy of the incident electrons\cite{Beenakker1}.
For the proximitized graphene here, both RAR and SAR can be realized
simultaneously for the fixed incident energy
since the presence of the spin-valley dependent
effective chemical potential $\mu_{eff}$
caused by the intrinsic spin-orbit coupling,
which is the origin of the spin-valley dependent double Andreev reflections.
Next, we present the numerical results for the reflection probabilities
and conductance with the specific chemical potential.

\begin{figure}[!htb]
\centerline{\includegraphics[width=0.9\columnwidth]{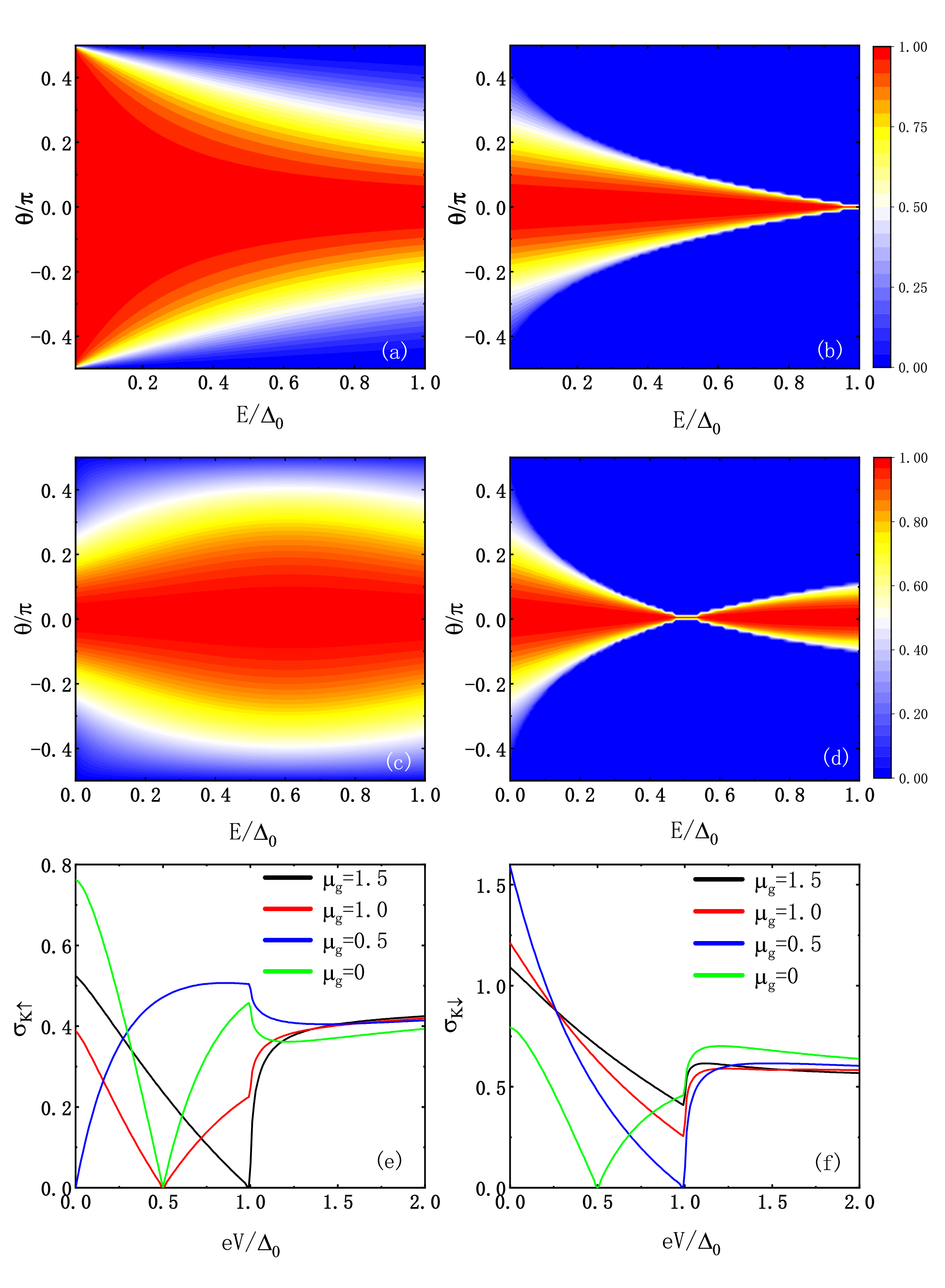}}
\caption{(a-d) The probabilities of the Andreev reflections in the space of ($E,\theta$)
for the injection of (a,c) the $K\uparrow$ electrons and (b,d) the $K\downarrow$ electrons
with the chemical potential $\mu_{g}=0.5\Delta_{0}$ (a,b) and $\mu_{g}=0$ (c,d).
When $\mu_{g}=0.5\Delta_{0}$, only the SAR happens in (a) for the $K\uparrow$ electrons
and only the RAR happens in (b) for the $K\downarrow$ electrons.
(e) and (f)
The normalized conductance spectra for (e) the $K\uparrow$ electrons and (f) the $K\downarrow$ electrons with different chemical potentials. The parameters $\Delta=0$ and $\lambda=0.5\Delta_{0}$ have been taken.}\label{fig2}
\end{figure}
For $\mu_{g}=0.5\Delta_{0}$, the linear dispersions are schematically shown in Figs. {\ref{fig1}}(c) and (d). The effective chemical potential $\mu_{eff}$ for the $K\uparrow$ electrons is $0$ and that for the $K\downarrow$ electrons is $\Delta_{0}$. The Dirac point for the $K\uparrow$ electrons is located at $E=0$ and that for the $K\downarrow$ electrons will be located at $E/\Delta_{0}=-1$. The pure SAR happens for the $K\uparrow$ incident electrons while the pure RAR happens for the $K\downarrow$ incident electrons in the superconducting gap as schematically shown in Figs. {\ref{fig1}}(a) and (b).
In Figs.{\ref{fig2}} (a) and (b), we plot the probabilities of SAR and RAR
for the $K\uparrow$ incident electrons and the $K\downarrow$ incident electrons,
respectively.
From the figures, one can find that for the definite chemical potential and a given incident energy $E$, RAR and SAR exist simultaneously but the incident angle range for the emergence of RAR is smaller than that for SAR.
When the energy $E$ increases, this character becomes more prominent.
This is because the actual RAR happens in the angle range $\vert \theta\vert<\arcsin[(\lambda+\mu_g-E)/(\lambda+\mu_g+E)]$ due to the conservation of the momentum along the $y$ axis in the reflection process.
Beyond this angle range, the wave function for the retro-Andreev reflected hole is attenuated due to its virtual wave vector.
It is obvious that the increasing energy $E$ will shrink the angle range.
Especially, as $E$ tends to $\Delta_{0}$, the angle range for RAR will tend to $0$ as shown in Fig. {\ref{fig2}}(b).
However, the occurrence of SAR does not have this angle limitation.
The above different properties of SAR and RAR can be well reflected in conductance in Figs. {\ref{fig2}} (e) and (f) (see the blue lines).
The subgap value of $\sigma_{K\downarrow}$ in Fig.{\ref{fig2}}(f) for the incident $K\downarrow$ electrons declines rapidly as the bias is raised.
The value will turn into zero for $eV/\Delta_{0}=1$.
However, the conductance $\sigma_{K\uparrow}$ in Fig.{\ref{fig2}}(e)
in the gap increases as $eV$ is raised.
Its peak value can be obtained for $eV/\Delta_{0}=1$.
Note, $\sigma_{K\uparrow}$ will tend to zero when $eV\rightarrow0$
although SAR is remarkable around $E=0$.
This is because the mode number $N_{\uparrow}(eV)$ will become zero
when $eV=0$ for $\lambda=\mu_{g}=0.5\Delta_{0}$ and $\Delta=0$.

In order to clarify the influences of the chemical potential
on the double Andreev reflections and conductance, the situations of $\mu_{g}=0$, $\Delta_{0}$ and $1.5\Delta_{0}$ are also considered. In these situations, the expressions for the wave functions and the reflection probabilities are given in APPENDIX. For $\mu_{g}=0$, the effective chemical potential $\mu_{eff}$
for the $K\uparrow$ electrons is $-\lambda = -0.5\Delta_{0}$
and that for the $K\downarrow$ electrons is $\lambda = 0.5\Delta_{0}$.
The conversion from RAR to SAR can be realized by changing the energy of the incident electrons, similar to the case of the isolated graphene\cite{Beenakker1}.
The energy range for RAR or SAR is the same for the $K\uparrow$ electrons and the $K\downarrow$ electrons when $\mu_{g}=0$.
RAR happens in the energy range $0<E/\Delta_{0}<0.5$ and SAR happens in the range $E/\Delta_{0}>0.5$.
Figs.{\ref{fig2}}(c) and (d) show the probabilities of the Andreev reflections
in the space of ($E,\theta$) for $\mu_g=0$.
When the change between SAR and RAR at $E/\Delta_{0}=0.5$ occurs,
the incident angle range for the occurrence of the Andreev reflection
reaches its minimum (maximum)
for the $K\downarrow$ ($K\uparrow$) incident electrons.
In addition, both the conductances $\sigma_{K\uparrow}$ and $\sigma_{K\downarrow}$
have their zero point at the change point between SAR and RAR
with $eV/\Delta_{0}=0.5$ as shown in Figs. {\ref{fig2}} (e) and (f) (see green lines).
However, for the $K\uparrow$ electrons responsible for RAR,
the Andreev reflected holes come from the $K'\downarrow$ valence band,
while the Andreev reflected holes come from the $K'\uparrow$
conduction band for the $K\downarrow$ electrons.
The zero point in $\sigma_{K\uparrow}$ originates from the vanishing mode number $N_{\uparrow}(eV)$ of the incident electron at $eV/\Delta_{0}=0.5$
while that in $\sigma_{K\downarrow}$ originates
from the vanishing Andreev reflected hole modes.

For $\mu_{g}=\Delta_{0}$, the effective chemical potential for the $K\uparrow$ electrons is $0.5\Delta_{0}$ and that for the $K\downarrow$ electrons is $1.5\Delta_{0}$.
Only RAR can emerge for the $K\downarrow$ electrons in the superconducting gap. For the $K\uparrow$ electrons, RAR emerges in the range $0<E/\Delta_{0}<0.5$ and SAR appears in the range $0.5<E/\Delta_{0}<1$.
The conductances for the $K\uparrow$ and the $K\downarrow$ electrons are presented in Figs.{\ref{fig2}}(e) and (f) (see red lines).
For both $K\uparrow$ and $K\downarrow$ indices,
the Andreev reflected holes come from their conduction band for RAR.
However, for the $K\uparrow$ electrons, the conversion from RAR to SAR in the superconducting gap can be realized when the energy rises across $E/\Delta_{0}=0.5$.
The conductance $\sigma_{K\uparrow}$ in Fig.{\ref{fig2}}(e)
has the character of the isolated graphene,
which possesses the zero point at the effective chemical potential, i.e., $eV/\Delta_{0}=(\mu_g-\lambda)/\Delta_{0}=0.5$,
due to the vanishing Andreev reflected hole modes.
For $\mu_{g}=1.5\Delta_{0}$, the effective chemical potential
for the $K\uparrow$ electrons is $\mu_g-\lambda=\Delta_{0}$ and that for the $K\downarrow$ electrons is $\mu_g+\lambda=2\Delta_{0}$.
Only RAR can happen in the superconducting gap for both the $K\uparrow$
and the $K\downarrow$ electrons.
The conductances are shown as the black lines in Figs.{\ref{fig2}}(e) and (f).
For the $K\uparrow$ electrons, the conductance has a zero point at $eV/\Delta_{0}=1$ due to the vanishing hole modes at $E/\Delta_{0}=1$.

From the above discussions on the numerical results, we can summarize the condition for the emergence of the specific Andreev reflection in the superconducting gap related to the electrons with the fixed spin-valley index according to the value of the effective chemical potential $\mu_{eff}$.
For $K\uparrow$ and $K\downarrow$ electrons, the effective chemical potential
$\mu_{eff}$ is equal to $\mu_g-\lambda$ and $\mu_g+\lambda$, respectively.
If $\mu_{eff}=0$, only SAR happens for $0<E<\Delta_{0}$.
The incident electrons for SAR and the reflected holes belong to the conduction band and the valence band, respectively.
If $0<\mu_{eff}<\Delta_{0}$, RAR happens in the energy range $0<E<\mu_{eff}$
and SAR happens in the range $\mu_{eff}<E<\Delta_{0}$.
For RAR, both the incident electrons and the reflected holes belong to the conduction bands. But for SAR, the electrons and holes respectively belong to the conduction band and the valence band.
If $\mu_{eff}\ge \Delta_{0}$, only RAR happens in the range $0<E<\Delta_{0}$
with both the incident electrons and the reflected holes in the conduction bands.
On the other hand, if $-\Delta_{0}<\mu_{eff}<0$, RAR happens in the range $0<E<-\mu_{eff}$ and SAR happens in the range $-\mu_{eff}<E<\Delta_{0}$.
If $\mu_{eff}\le-\Delta_{0}$, only RAR can happen in the range $0<E<\Delta_{0}$.
These conditions apply to both the $K\uparrow$ and the $K\downarrow$ electrons which have $\mu_{eff}=\mu_{g}-\lambda$ and $\mu_{eff}=\mu_{g}+\lambda$.
The different effective chemical potential can bring about the different types of Andreev reflections for the electrons with the different spin-valley index. Furthermore, the spin-valley dependent Andreev reflections coexist in the proximitized graphene. This is distinct from the isolated graphene without the intrinsic spin-orbit coupling.

\begin{figure}[!htb]
\centerline{\includegraphics[width=0.9\columnwidth]{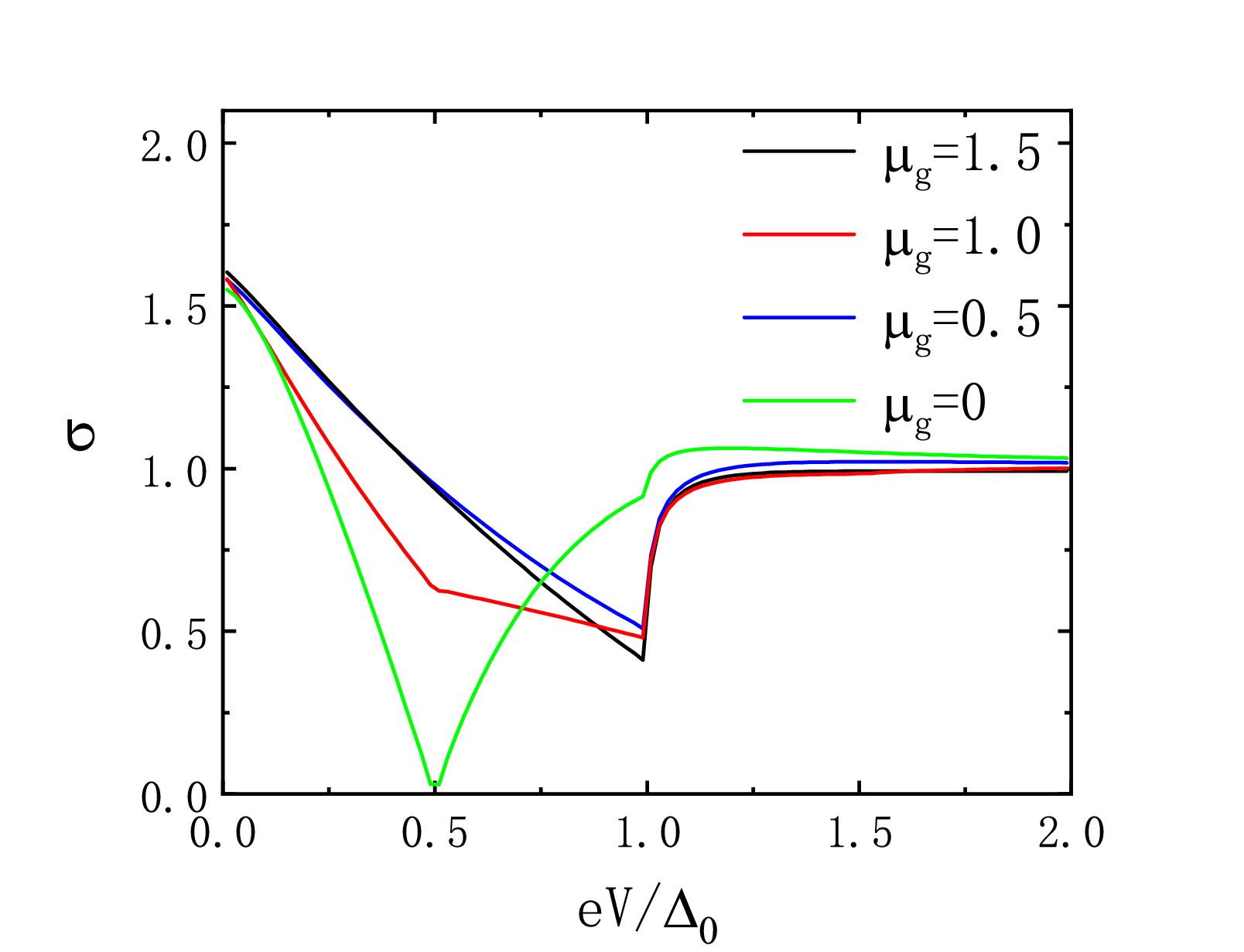}}
\caption{The normalized total conductance spectra for different chemical potentials with $\Delta=0$ and $\lambda=0.5\Delta_{0}$.}\label{fig3}
\end{figure}
Next, we give the normalized total conductance in Fig.{\ref{fig3}},
which is contributed from both the $K\uparrow$ electrons and the $K\downarrow$ electrons. Due to the emergence of the spin-valley dependent Andreev reflections,
the conductance spectra for the proximitized graphene/superconductor junction here are very different from those for the isolated graphene/superconductor junction\cite{Beenakker1}.
For the isolated graphene, the conductance has the zero point for $\mu_{g}<\Delta_0$.
But for the proximitized graphene, except for $\mu_{g}=0$,
the conductance can not reach zero.
The conductance spectra here are also different from those based on the double Andreev reflections in the three-dimensional semimetals\cite{Hou,QCheng3}.
The conductance spectra in Fig.{\ref{fig3}} include
the information of the intrinsic spin-orbit coupling $\lambda$.
For example, for $\mu_{g}=0$ (the green line),
the position for the zero point is just the value of $\lambda$.
For $\mu_{g}=\Delta_{0}$ (the red line), the position for the turning point
of the curve is just $\mu_{g}-\lambda$.
Therefore, our results provide an effective method
based on the transport experiment for the detection of the intrinsic spin-orbit coupling strength in the proximitized graphene system.
\subsection{$\Delta\ne 0$}
Now, we turn to the graphene having the finite pseudospin
staggered potential with $\Delta=0.2\Delta_{0}$.
The effects of the finite staggered potential are to open a gap of $2\Delta$
in the linear dispersions of electrons and holes and to modify the mode numbers $N_{\uparrow}(eV)$ and $N_{\downarrow}(eV)$.
The gap for the $K\uparrow$ $(K\downarrow)$ electrons spans
from $-\Delta+(-)\lambda-\mu_{g}$ to $\Delta+(-)\lambda-\mu_{g}$ and the gap for the associated holes spans from $-\Delta-(+)\lambda+\mu_{g}$ to $\Delta-(+)\lambda+\mu_{g}$.
If the gap is situated in the superconducting gap $-\Delta_{0}<E<\Delta_{0}$,
the Andreev reflection and then the conductance will be significantly modified. If the gap is situated in $|E|>\Delta_{0}$,
the Andreev reflection and the conductance almost keep unchanged.
Here the condition for the emergence of Andreev reflections summarized
in Sec.\uppercase\expandafter{\romannumeral 3}A holds
for the situation of the finite potential.

\begin{figure}[!htb]
\centerline{\includegraphics[width=0.9\columnwidth]{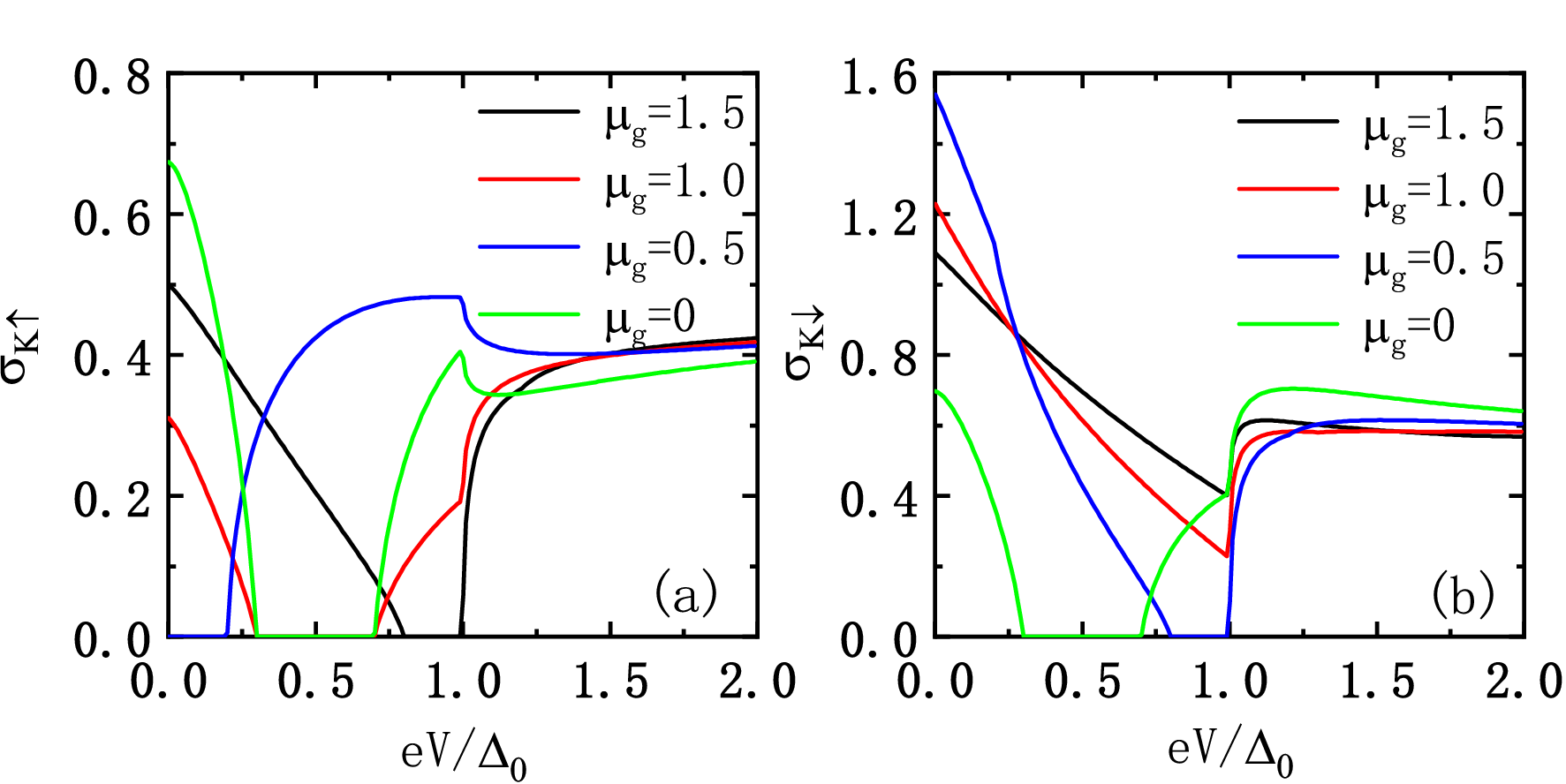}}
\caption{The normalized conductance spectra with different chemical potentials for (a) the $K\uparrow$ electrons and (b) the $K\downarrow$ electrons at $\Delta=0.2\Delta_{0}$ and $\lambda=0.5\Delta_{0}$.}
\label{fig4}
\end{figure}

We still first take $\mu_{g}=0.5\Delta_{0}$ as an example.
When the staggered potential is introduced in graphene,
the energy spectrum opens a gap.
Here the dispersions are shown schematically in Figs. {\ref{fig1}}(e) and (f).
There is no injection of the $K\uparrow$ electrons
in the energy gap $-0.2\Delta_{0}<E<0.2\Delta_{0}$.
As a result, the conductance $\sigma_{K\uparrow}$ is zero in $0<eV/\Delta_{0}<0.2$ as shown in Fig.{\ref{fig4}}(a) (see the blue line).
For the $K'\uparrow$ holes associated with the $K\downarrow$ electrons,
the gap spans from $E=0.8\Delta_{0}$ to $E=1.2\Delta_{0}$ in which the holes are absent. In this energy range, there is no Andreev reflection for the incident $K\downarrow$ electrons.
On the other hand, there is also no tunneling of the $K\downarrow$ electrons in the range $E<\Delta_{0}$ because of the presence of the superconducting gap. Hence, the conductance $\sigma_{K\downarrow}$ for the $K\downarrow$ electrons is zero in $0.8<eV/\Delta_{0}<1$ as shown in Fig.{\ref{fig4}} (b) (see the blue line). While $eV/\Delta_{0}>1$, the normal
tunneling occurs and then $\sigma_{K\downarrow}$ is non-zero.

For $\mu_{g}=0$, the gaps for both the $K\uparrow$ electrons
and the $K'\uparrow$ holes associated with the $K\downarrow$ electrons span from $0.3\Delta_{0}$ to $0.7\Delta_{0}$. Accordingly, the conductances $\sigma_{K\uparrow}$ and $\sigma_{K\downarrow}$ become zero in $0.3<eV/\Delta_{0}<0.7$ as shown in Figs.{\ref{fig4}}(a) and (b) (see green lines).
For $\mu_{g}=\Delta_{0}$, the gap spanning from $0.3\Delta_{0}$ to $0.7\Delta_{0}$ for the holes associated with the $K\uparrow$ electrons causes the zero conductance of $\sigma_{K\uparrow}$ in $0.3<eV/\Delta_{0}<0.7$ as shown in Fig.{\ref{fig4}}(a) (see the red line).
For $\mu_{g}=1.5\Delta_{0}$, the gap for the holes associated with the $K\uparrow$ electrons spans from $0.8\Delta_{0}$ to $1.2\Delta_{0}$, which leads to the zero conductance of $\sigma_{K\uparrow}$ in $0.8<eV/\Delta_{0}<1$ as shown in Fig.{\ref{fig4}}(a) (see the black line).
On the other hand, if $\mu_{g}>\Delta_0-\lambda+\Delta=0.7\Delta_0$,
the gaps of both the $K\downarrow$ electrons and their associated holes locate
the outer of the superconducting gap.
As a result, the conductance $\sigma_{K\downarrow}$ is always non-zero
for $eV$ within the superconducting gap (see the red and black curves
in Fig.{\ref{fig4}}(b)).

Fig.{\ref{fig5} gives the normalized total conductances for different chemical potentials with $\Delta=0.2\Delta_{0}$.
From the conductance spectra, one can also extract the information
about the values of the pseudospin staggered potential and the intrinsic spin-orbit coupling strength.
For example, the first turning point of the conductance
for $\mu_{g}=0.5\Delta_{0}$ happens at $eV=\Delta+\lambda-\mu_g$
while the second turning point emerges at $eV=-\Delta+\lambda+\mu_{g}$.
For $\mu_{g}=0$, the zero conductance can be observed in the range $-\Delta+\lambda<eV<\Delta+\lambda$. For $\mu_{g}=\Delta_{0}$, the two turning points of the conductance curve happen at $eV=-\Delta-\lambda+\mu_{g}$ and $\Delta-\lambda+\mu_{g}$, respectively. Therefore, the measurement of the transport properties of the proximitized graphene/superconductor junction
can help determine the two important parameters induced in graphene.

\begin{figure}[!htb]
\centerline{\includegraphics[width=0.9\columnwidth]{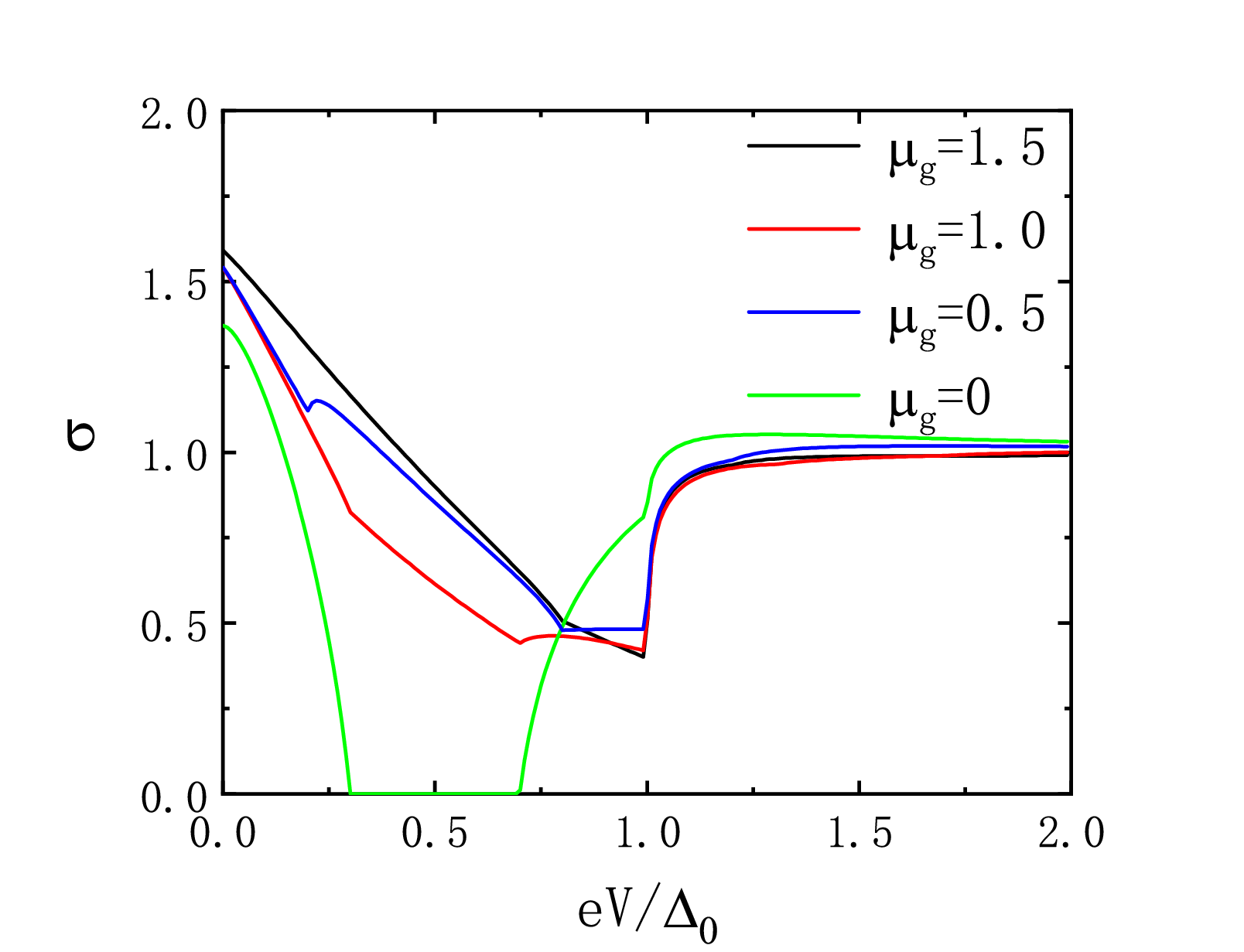}}
\caption{The normalized total conductance spectra for different chemical potentials with $\Delta=0.2\Delta_{0}$ and $\lambda=0.5\Delta_{0}$.}\label{fig5}
\end{figure}
\subsection{Conductance in the $(\mu_{g},eV)$ space}
\begin{figure}[!htb]
\centerline{\includegraphics[width=1\columnwidth]{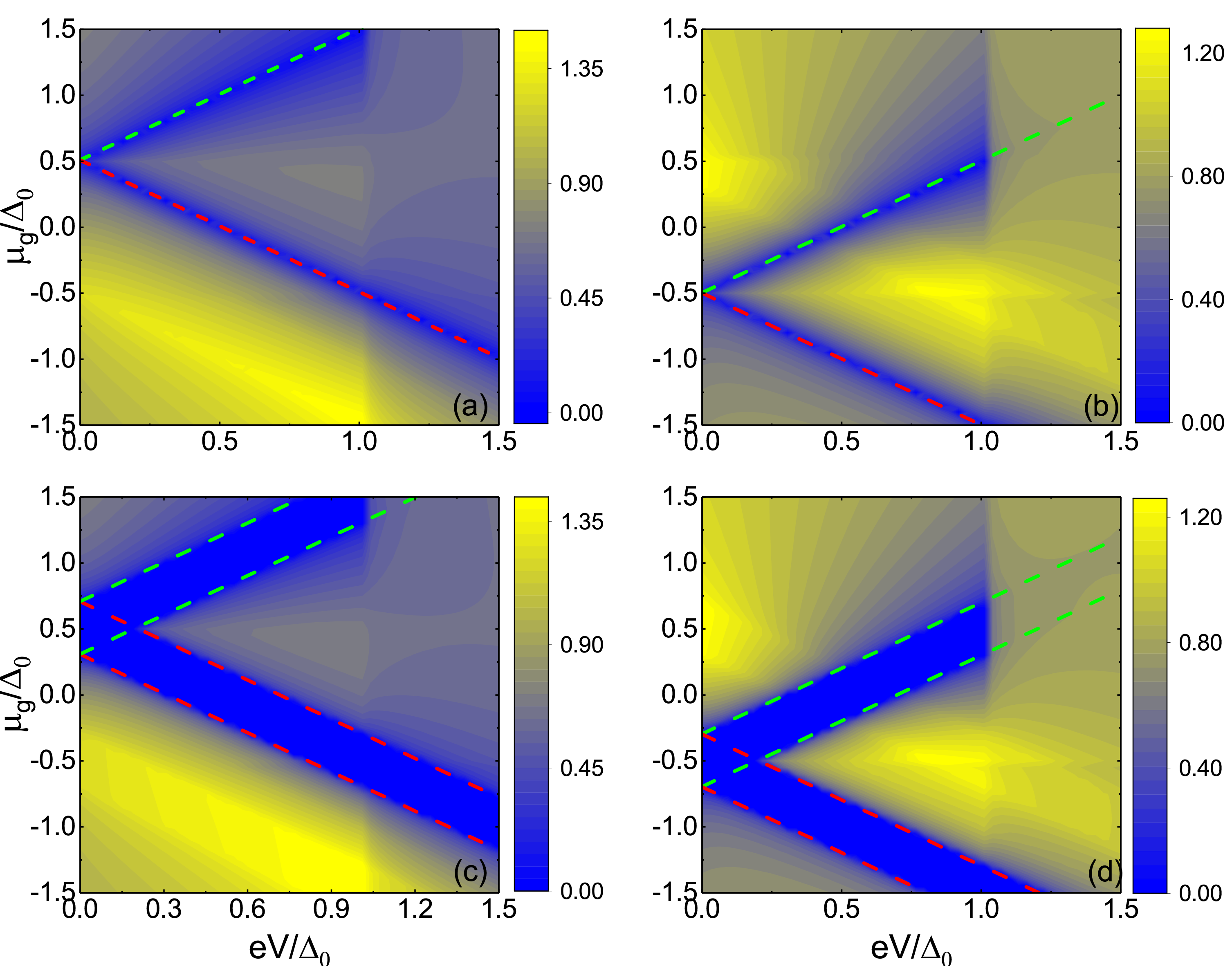}}
\caption{The conductances (a) $\sigma_{K\uparrow}$ and (b) $\sigma_{K\downarrow}$ in the $(\mu_{g},eV)$ space for $\Delta=0$.
The conductances (c) $\sigma_{K\uparrow}$ and (d) $\sigma_{K\downarrow}$
in the $(\mu_{g},eV)$ space for $\Delta=0.2\Delta_{0}$.
The parameter $\lambda=0.5\Delta_{0}$ has been taken.
The dashed lines in (a) and (b) denote the positions for the zero conductance. The dashed lines in (c) and (d) denote the boundary lines of the zones for the zero conductance.
For clarity, the linear values on the scale bars in (a-d) are set
to be $\sqrt{\sigma_{K\uparrow}}$ or $\sqrt{\sigma_{K\downarrow}}$.
}\label{fig6}
\end{figure}
To get a more complete understanding of the dependence of Andreev reflections
on the chemical potential, we present the conductance in the $(\mu_{g},eV)$ space. We first consider the conductances $\sigma_{K\uparrow}$ and $\sigma_{K\downarrow}$ for the case of $\lambda=0.5\Delta_{0}$ in Fig.{\ref{fig6}}. Furthermore, we take $\Delta=0$ in Figs.{\ref{fig6}}(a) and (b) and $\Delta=0.2\Delta_{0}$ in Figs.{\ref{fig6}}(c) and (d). In Fig.{\ref{fig6}}(a), the conductance for the injection of the $K\uparrow$ electrons is plotted. The positions of the zero points for the conductance are denoted by two dashed straight lines. The red one is determined by the relation $\mu_{g}=\lambda-eV$ and the green one is determined by $\mu_{g}=\lambda+eV$.
The zero points along the red line correspond to the Dirac point of the $K\uparrow$ electrons where the mode number for the electrons is vanished.
The zero points along the green line correspond to the Dirac point of the $K'\downarrow$ holes
where the mode number for the reflected holes is vanished.
In the region below the red line, RAR happens in the superconducting gap. Both the $K\uparrow$ electrons and the Andreev reflected $K'\downarrow$ holes are from their valence bands. In the region above the green line, the Andreev reflection is also of the retro-type. However, both the $K\uparrow$ electrons and the $K'\downarrow$ holes are from their conduction bands. In the region in between the green line and the red line, SAR happens in the superconducting gap. The $K\uparrow$ electrons come from the conduction band while the $K'\downarrow$ holes come from the valence band.

In Fig.{\ref{fig6}}(b), the conductance for the injection of the $K\downarrow$ electrons is plotted. The green and the red dashed lines denoting the positions of the zero conductance are determined by $\mu_{g}=-\lambda+eV$ and $\mu_{g}=-\lambda-eV$, respectively. The types of the Andreev reflection happening in the three regions divided by the two dashed lines are the same with those in Fig.{\ref{fig6}}(a) for the $K\uparrow$ electrons. But, the dashed lines in Fig.{\ref{fig6}}(a) and (b) do not coincide. As a result, the spin-valley dependent double Andreev reflections will be expected in the region with the boundary lines $\mu_{g}=\lambda\pm eV$ and $\mu_{g}=-\lambda+eV$ and the region with the boundary lines $\mu_{g}=-\lambda\pm eV$ and $\mu_{g}=\lambda-eV$. In the former (latter) region, SAR happens for the $K\uparrow(K\downarrow)$ electrons and RAR happens for the $K\downarrow(K\uparrow)$ electrons. For $\Delta=0.2\Delta_{0}$, the dashed lines for the zero conductance will expand to finite zones for the zero conductance due to the open of gaps at the Dirac points of electrons and holes as shown in Figs.{\ref{fig6}}(c) and (d). For the $K\uparrow(K\downarrow)$ electrons, the green line expands to the zone between $\mu_{g}=+(-)\lambda+eV-\Delta$ and $\mu_{g}=+(-)\lambda+eV+\Delta$ while the red line expands to the zone between $\mu_{g}=+(-)\lambda-eV-\Delta$ and $\mu_{g}=+(-)\lambda-eV+\Delta$. The boundary lines of these zones are plotted as the red and green dashed lines in Figs.{\ref{fig6}}(c) and (d).

\begin{figure}[!htb]
\centerline{\includegraphics[width=1\columnwidth]{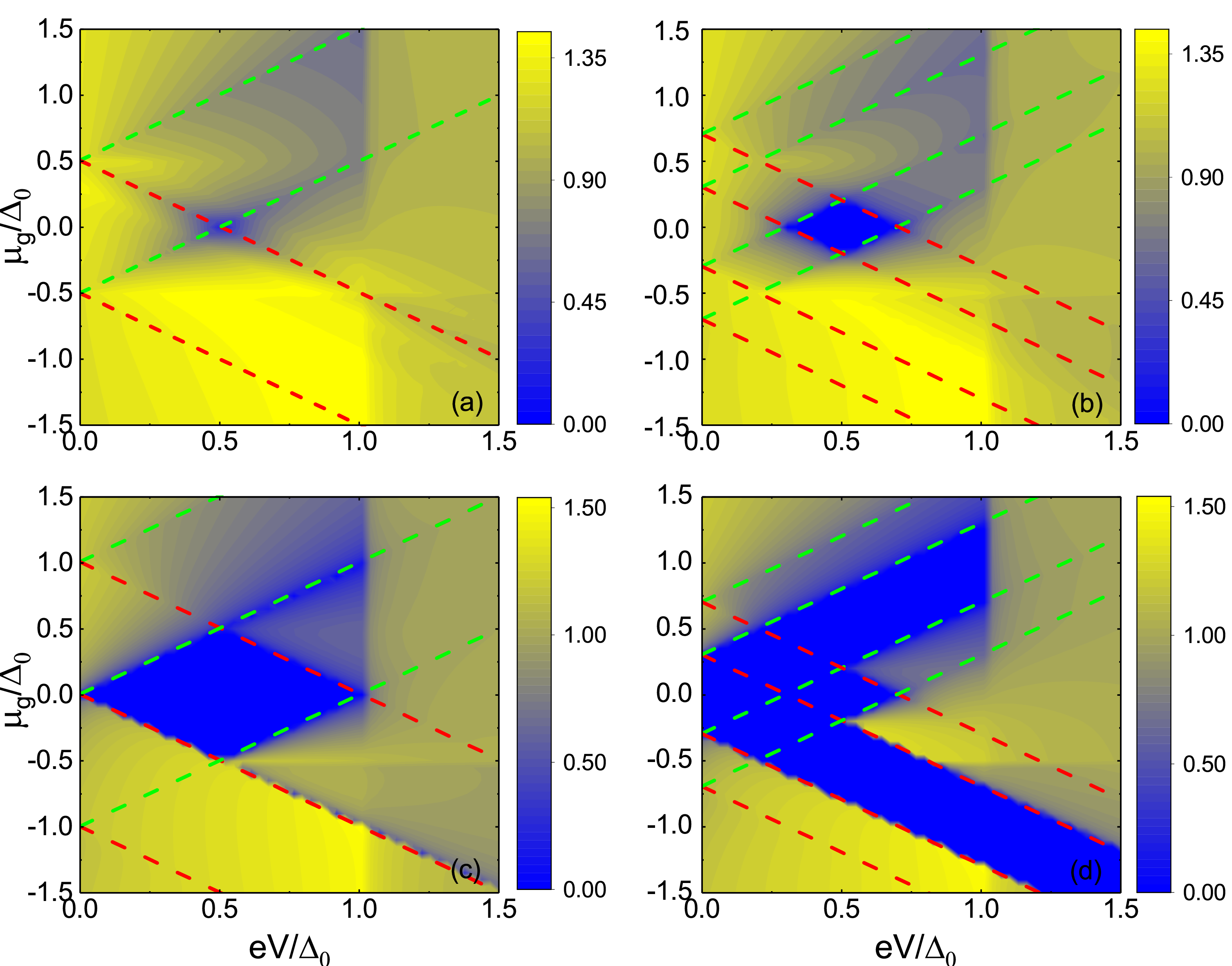}}
\caption{The normalized total conductance $\sigma$ in the $(\mu_{g},eV)$ space for (a)$\lambda=0.5\Delta_{0},\Delta=0$, (b)$\lambda=0.5\Delta_{0},\Delta=0.2\Delta_{0}$, (c)$\lambda=0.5\Delta_{0},\Delta=0.5\Delta_{0}$ and (d)$\lambda=0.2\Delta_{0},\Delta=0.5\Delta_{0}$.
The red and green dashed lines in (a) denote the positions for the zero points of $\sigma_{K\uparrow}$ and $\sigma_{K\downarrow}$. The dashed lines in (b)-(d) denote the boundary lines of the zones for the zero conductances of $\sigma_{K\uparrow}$ and $\sigma_{K\downarrow}$.
For clarity, the linear values on the scale bars in (a-d) are set
to be $\sqrt{\sigma}$.
}\label{fig7}
\end{figure}
The normalized total conductances in the $(\mu_{g},eV)$ space for $(\lambda,\Delta)=(0.5\Delta_{0},0)$ and $(\lambda,\Delta)=(0.5\Delta_{0},0.2\Delta_{0})$ are presented in Figs.{\ref{fig7}}(a) and (b), respectively.
The conductance in Fig.{\ref{fig7}}(a) is the sum of $\sigma_{K\uparrow}$ and $\sigma_{K\downarrow}$ in Figs.{\ref{fig6}}(a) and (b). The conductance in Fig.{\ref{fig7}}(b) is the sum of $\sigma_{K\uparrow}$ and $\sigma_{K\downarrow}$ in Figs.{\ref{fig6}}(c) and (d).
For $\mu_{g}/\Delta_{0}<-\lambda/\Delta_{0}=-0.5$, the total conductance increases when $eV$ is raised in the superconducting gap. The conductance $\sigma_{K\uparrow}$ possesses large value due to its large mode number, which will eliminate the effect of the zero points of $\sigma_{K\downarrow}$.
On the other hand, for $\mu_{g}/\Delta_{0}>-\lambda/\Delta_{0}=-0.5$,
the emergences of these zero or turning points
become possible which provide the information of the magnitude of the
staggered potential $\Delta$ and the intrinsic spin-orbit coupling $\lambda$.
The condition for the emergence of the zero total conductance is
the simultaneously vanishing $\sigma_{K\uparrow}$ and $\sigma_{K\downarrow}$,
which corresponds to the cross point of the dashed lines
in Fig.{\ref{fig7}}(a) for $\Delta=0$
or the cross zone (the parallelogram) surrounded by the green and red
dashed lines in Fig.{\ref{fig7}}(b) for $\Delta=0.2\Delta_{0}$.
For example, the cross point of the red dashed line and the green dashed line
at $(\mu_{g},eV)= (0,0.5\Delta_0)$ in Fig.{\ref{fig7}}(a) corresponds to the zero point of the total conductance in Fig.{\ref{fig3}}.
Then we can obtain the staggered potential and spin-orbit coupling
$(\lambda,\Delta)=(0.5\Delta_0,0)$, straightforwardly.
In Fig.{\ref{fig7}}(b), the cross parallelogram zone
with the total conductance $\sigma=0$ is clearly visible
and its four vertices are $(\mu_{g}/\Delta_0,eV/\Delta_0)
=(0,0.3)$, $(0.2,0.5)$, $(0,0.7)$, and $(-0.2,0.5)$.
From this parallelogram, one can easily obtain $\lambda$ and $\Delta$.
In addition, the total conductance $\sigma$ also possesses
some turning points (e.g. see the green dashed lines in the region
$\mu_g/\Delta_0>0.5$), which are caused by the vanishing of one of the two conductances
$\sigma_{K\uparrow}$ and $\sigma_{K\downarrow}$.

If we keep $\lambda=0.5\Delta_{0}$ and increase $\Delta$ from $0.2\Delta_{0}$
to $0.5\Delta_{0}$, the two green and two red dashed lines in the middle in Fig.{\ref{fig7}}(b) will repectively merge into the middle green line and the middle red line in Fig.{\ref{fig7}}(c).
The condition for the formation of the zero points
and the turning points for the total conductance is the same with that for $(\lambda,\Delta)=(0.5\Delta_{0},0.2\Delta_{0})$ in Fig.{\ref{fig7}}(b).
The area of the cross zone (the parallelogram) with
the total conductance $\sigma=0$ will become lager.
If we keep $\Delta=0.5\Delta_{0}$ and decrease $\lambda$ from $0.5\Delta_{0}$ to $0.2\Delta_{0}$, the two green (red) dashed lines in the middle in Fig.{\ref{fig7}}(b) will exchange their positions as shown in Fig.{\ref{fig7}}(d).
As a result, the total conductance will be zero in the zone between
the green dashed lines and the red dashed lines in the middle
as shown in Fig.{\ref{fig7}}(d).
In all situations, the staggered potential and the intrinsic spin-orbit coupling can be detected by the transport measurement in a larger range of the chemical potential.

Finally, we give a short discussion on the influence of
the Rashba spin-orbit coupling on our numerical results.
The Rashba spin-orbit coupling is not included in our model
for the proximitized graphene.
Generally, the introduction of the Rashba spin-orbit coupling will generate two effects. The first is to open a gap in the linear dispersions of graphene. The second is to mix the up spin and the down spin and spin will not be a good quantum number for electrons and holes.
Nevertheless, the types of the Andreev reflections are mainly determined by the electronic structure\cite{QCheng4}. From the electronic structure of the proximitized graphene\cite{Gmitra,Wang,Zubair,Cysne}, it is believed that the double Andreev reflections can still survive in the presence of the Rashba spin-orbit coupling.

As for the experimental feasibility of our junction, recent experiments in Refs.[\onlinecite{Zihlmann,Wakamura,Wang}] demonstrate that the intrinsic spin-orbit coupling can be induced in graphene by the transition metal dichalcogenide substrates in a wide range of the gate voltage from $-40V$ to $8V$. The intrinsic spin-orbit coupling can also be induced in graphene on metal substrates\cite{Frank}, which provides more experimental scheme for the realization of our junction. Furthermore, the type of the Andreev reflection and the boundary of SAR and RAR can be well detected as long as charge puddles in graphene are less than two orders of magnitude of the superconducting gap\cite{Cheng2}. The features of the Andreev reflections and the conductance in our junction are also insensitive to the large doping in superconductor especially when the superconductor is heavily doped\cite{Beenakker1}.

\section{\label{sec4}CONCLUSIONS}
We study the Andreev reflections in the proximitized graphene/supercondcutor junction with the induced pseudospin staggered potential and the intrinsic spin-orbit coupling.
The quadruple degeneracy of spin and valley in the isolated graphene is broken in the proximitized one due to the presence of the staggered potential
and the spin-orbit coupling, which brings about the spin-valley dependent energy bands of electrons and holes.
The coexistence of RAR and SAR, which are also spin-valley dependent,
can be realized in the junction.
The strong dependences of the spin-valley resolved double Andreev reflections on the chemical potential are investigated in details.
The anomalous conductance spectra distinct from those for the isolated graphene/superconductor junction are presented, which provide
an experimentally feasible way to detect the strength of the
staggered potential and the intrinsic spin-orbit coupling.
Our numerical results laid the foundation for the establishment and the development of the spin-valley electronics based on superconductor.

\section*{\label{sec5}ACKNOWLEDGMENTS}
This work was financially supported by
NSF-China under Grants Nos. 11921005 and 11447175,
the Innovation Program for Quantum Science and Technology (2021ZD0302403),
the Strategic Priority Research Program of Chinese Academy of Sciences (XDB28000000)
and the Natural Science Foundation of Shandong Province under Grants No. ZR2017QA009.

\section*{APPENDIX}
\setcounter{equation}{0}
\renewcommand{\theequation}{A.\arabic{equation}}
Here, we present the wave functions and the reflection probabilities for different values of $\lambda$ and $\mu_{g}$ in graphene. Firstly, we consider the injection of a $K\uparrow$ electron. For $\lambda-\mu_{g}\ge0$, the wave function in graphene can be solved as
\begin{eqnarray}
\begin{split}
\psi_{G}^{K\uparrow}&=\left(\begin{array}{c}
-\chi_{1}\eta_{e\uparrow-} \\
1 \\
0\\
0\end{array}\right)e^{-i k_{x}^{e\uparrow-}x}
+r_{1}\left(\begin{array}{c}
\chi_{1}\eta_{e\uparrow-}^{*}\\
1\\
0\\
0\end{array}\right)e^{i k_{x}^{e\uparrow-}x}\\
&+
r_{a1}\left(\begin{array}{c}
0\\
0\\
\chi_{2}\eta_{h\downarrow+}\\
1\end{array}\right)e^{-i k_{x}^{h\downarrow+}x},\label{Awfku01}
\end{split}
\end{eqnarray}
for $0< E<\lambda-\mu_{g}$. The expressions of the symbols are $\eta_{e\uparrow-}=(k_{x}^{e\uparrow-}+i k_{y})/k_{e\uparrow-}$, $k_{x}^{e\uparrow-}=\sqrt{[E-\lambda+\mu_{g}]^2-\Delta^2-\hbar^2v_{F}^2k_{y}^2}/{\hbar v_{F}}$ and $k_{e\uparrow-}=\sqrt{{k_{x}^{e\uparrow-}}^2+k_{y}^2}$. The probability of RAR in the energy range $0< E<\lambda-\mu_{g}$ is given by
\begin{eqnarray}
R_{a1}=-\frac{\chi_{2}k_{x}^{h\downarrow+}k_{e\uparrow-}}{\chi_{1}k_{x}^{e\uparrow-}k_{h\downarrow+}}\vert r_{a1}\vert^2.\label{ARa11}
\end{eqnarray}
For $E>\lambda-\mu_{g}$, the wave function is the same with that in Eq.(\ref{wfku})
and the probability of SAR is given by Eq.(\ref{Ra1}).

For $\lambda-\mu_{g}<0$, the wave function in graphene can be solved as
\begin{eqnarray}
\begin{split}
\psi_{G}^{K\uparrow}&=\left(\begin{array}{c}
\chi_{1}\eta_{e\uparrow+} \\
1 \\
0\\
0\end{array}\right)e^{i k_{x}^{e\uparrow+}x}
+r_{1}\left(\begin{array}{c}
-\chi_{1}\eta_{e\uparrow+}^{*}\\
1\\
0\\
0\end{array}\right)e^{-i k_{x}^{e\uparrow+}x}\\
&+
r_{a1}\left(\begin{array}{c}
0\\
0\\
-\chi_{2}\eta_{h\downarrow-}\\
1\end{array}\right)e^{i k_{x}^{h\downarrow-}x},\label{Awfkude1}
\end{split}
\end{eqnarray}
for $0< E<\mu_{g}-\lambda$ with $k_{x}^{h\downarrow-}=\sqrt{[E+\lambda-\mu_{g}]^2-\Delta^2-\hbar^2v_{F}^2k_{y}^2}/{\hbar v_{F}}$, $\eta_{h\downarrow-}=(k_{x}^{h\downarrow-}-i k_{y})/k_{h\downarrow-}$ and $k_{h\downarrow-}=\sqrt{{k_{x}^{h\downarrow-}}^2+k_{y}^2}$. The probability of RAR in the energy range $0< E<\mu_{g}-\lambda$ is given by
\begin{eqnarray}
R_{a1}=-\frac{\chi_{2}k_{x}^{h\downarrow-}k_{e\uparrow+}}{\chi_{1}k_{x}^{e\uparrow+}k_{h\downarrow-}}\vert r_{a1}\vert^2.\label{ARa13}
\end{eqnarray}
The wave function and the probability of SAR for $E>\mu_{g}-\lambda$ are the same with those in Eqs.(\ref{wfku}) and (\ref{Ra1}), respectively.

Secondly, we consider the injection of a $K\downarrow$ electron. For $\lambda+\mu_{g}\ge 0$, the wave functions are the same with those in Eqs.(\ref{wfkd1}) and (\ref{wfkd2}). The probabilities are the same with those in Eqs.(\ref{Ra21}) and (\ref{Ra22}). For $\lambda+\mu_{g}<0$, the wave function is given by
\begin{eqnarray}
\begin{split}
\psi_{G}^{K\downarrow}&=\left(\begin{array}{c}
-\chi_{3}\eta_{e\downarrow-}\\
1\\
0\\
0\end{array}\right)e^{-i k_{x}^{e\downarrow-}x}+r_{2}\left(\begin{array}{c}
\chi_{3}\eta_{e\downarrow-}^{*}\\
1\\
0\\
0\end{array}\right)e^{i k_{x}^{e\downarrow-}x}\\
&+r_{a2}\left(\begin{array}{c}
0\\
0\\
\chi_{4}\eta_{h\uparrow+}\\
1\end{array}\right)e^{-i k_{x}^{h\uparrow+}x},\label{Awfkd1}
\end{split}
\end{eqnarray}
for $0<E<\vert\lambda+\mu_{g}\vert$ with $k_{x}^{e\downarrow-}=\sqrt{[E+\lambda+\mu_{g}]^2-\Delta^2-\hbar^2v_{F}^2k_{y}^2}/\hbar v_{F}$, $\eta_{e\downarrow-}=(k_{x}^{e\downarrow-}+ik_{y})/k_{e\downarrow-}$ and $k_{e\downarrow-}=\sqrt{{k_{x}^{e\downarrow-}}^2+k_{y}^2}$. The probability of RAR in the range $0<E<\vert\lambda+\mu_{g}\vert$ is given by
\begin{eqnarray}
R_{a2}=-\frac{\chi_{4}k_{x}^{h\uparrow+}k_{e\downarrow-}}{\chi_{3}k_{x}^{e\downarrow-}k_{h\uparrow+}}\vert r_{a2}\vert^2.\label{ARa21}
\end{eqnarray}
For $E>\vert\lambda+\mu_{g}\vert$, the wave function is the same with that in Eq.(\ref{wfkd2}) and the probability of SAR is the same with that in Eq.(\ref{Ra22}).


\section*{REFERENCES}
\begin{thebibliography}{}
\bibitem{Andreev}
A. F. Andreev, Zh. Eksp. Teor. Fiz. \textbf{46}, 1823 (1964) [Sov. Phys. JETP 19, 1228 (1964)].
\bibitem{Tanaka1}
Y. Tanaka, and S. Kashiwaya, Phys. Rev. Lett. \textbf{74}, 3451 (1995).
\bibitem{Tanaka2}
Y. Tanaka, T. Kokkeler, and A. Golubov, Phys. Rev. B \textbf{105}, 214512 (2022).
\bibitem{Takabatake}
Y. Takabatake, S.-I. Suzuki, and Y. Tanaka, Phys. Rev. B \textbf{103}, 184515 (2021).
\bibitem{Anwar}
M. S. Anwar, S. L. Lee, R. Ishiguro, Y. Sugimoto, Y. Tano, S. J. Kang, Y. J. Shin, S. Yonezawa, D. Manske, H. Takayanagi, T. W. Noh, and Y. Maeno, Nat. Commun. \textbf{7}, 13220 (2016).
\bibitem{Zutic}
I. $\check{\text{Z}}$uti$\acute{\text{c}}$, J. Fabian, and S. D. Sarma, Rev. Mod. Phys. \textbf{76}, 323 (2004).
\bibitem{Zheng}
Z. Zheng, D. Y. Xing, G. Sun, and J. Dong, Phys. Rev. B \textbf{62}, 14326 (2000).
\bibitem{Hirai}
T. Hirai, Y. Tanaka, N. Yoshida, Y. Asano, J. Inoue, and S. Kashiwaya, Phys. Rev. B \textbf{67}, 174501 (2003).
\bibitem{Buzdin1}
A. I. Buzdin, Rev. Mod. Phys. \textbf{77}, 935 (2005).
\bibitem{Buzdin2}
A. Buzdin, Phys. Rev. Lett. \textbf{101}, 107005 (2008).
\bibitem{Goldobin}
E. Goldobin, D. Koelle, R. Kleiner, and R. G. Mints, Phys. Rev. Lett. \textbf{107}, 227001 (2011).
\bibitem{QCheng1}
Q. Cheng, and Q.-F. Sun, Phys. Rev. B \textbf{99}, 184507 (2019).
\bibitem{Linder1}
J. Linder, and J. W. A. Robinson, Nat. Phys. \textbf{11}, 307 (2015).
\bibitem{Beenakker1}
C. W. J. Beenakker, Phys. Rev. Lett. \textbf{97}, 067007 (2006).
\bibitem{Beenakker2}
C. W. J. Beenakker, Rev. Mod. Phys. \textbf{80}, 1337 (2008).
\bibitem{Linder2}
J. Linder, and A. Sudb$\o$, Phys. Rev. Lett. \textbf{99}, 147001 (2007).
\bibitem{Linder3}
J. Linder, A. M. Black-Schaffer, T. Yokoyama, S. Doniach, and A. Sudb$\o$, Phys. Rev. B \textbf{80}, 094522 (2009).
\bibitem{Cayssol}
J. Cayssol, Phys. Rev. Lett. \textbf{100}, 147001 (2008).
\bibitem{Zhang}
Q. Zhang, D. Fu, B. Wang, R. Zhang, and D. Y. Xing, Phys. Rev. Lett. \textbf{101}, 047005 (2008).
\bibitem{Schelter}
J. Schelter, B. Trauzettel, and P. Recher, Phys. Rev. Lett. \textbf{108}, 106603 (2012).
\bibitem{Cheng1}
S.-G. Cheng, Y. Xing, J. Wang, and Q.-F. Sun, Phys. Rev. Lett. \textbf{103}, 167003 (2009).
\bibitem{Xing}
Y. Xing, J. Wang, and Q.-F. Sun, Phys. Rev. B \textbf{83}, 205418 (2011).
\bibitem{Cheng2}
S.-G. Cheng, H. Zhang, and Q.-F. Sun, Phys. Rev. B \textbf{83}, 235403 (2011).

\bibitem{Majidi}
L. Majidi and M. Zareyan, Phys. Rev. B, \textbf{86}, 075443 (2012).

\bibitem{Zihlmann}
S. Zihlmann, A. W. Cummings, J. H. Garcia, M. Kedves, K. Watanabe, T. Taniguchi, C. Sch$\ddot{o}$nenberger, and P. Makk, Phys. Rev. B \textbf{97}, 075434 (2018).
\bibitem{Zollner1}
K. Zollner, and J. Fabian, Phys. Rev. Lett. \textbf{128}, 106401 (2022); Phys. Rev. B \textbf{106}, 035137 (2022).
\bibitem{Zollner2}
K. Zollner, A. W. Cummings, S. Roche, and J. Fabian, Phys. Rev. B \textbf{103}, 075129 (2021).
\bibitem{Khatibi}
Z. Khatibi, and S. R. Power, Phys. Rev. B \textbf{106}, 125417 (2022).
\bibitem{Wakamura}
T. Wakamura, F. Reale, P. Palczynski, M. Q. Zhao, A. T. C. Johnson, S. Gu$\acute{\text{e}}$ron, C. Mattevi, A. Ouerghi, and H. Bouchiat, Phys. Rev. B \textbf{99}, 245402, (2019).
\bibitem{Frank}
T. Frank, M. Gmitra, and J. Fabian, Phys. Rev. B \textbf{93}, 155142 (2016).
\bibitem{Gmitra}
M. Gmitra, and J. Fabian, Phys. Rev. B \textbf{92}, 155403 (2015).
\bibitem{Wang}
Z. Wang, D.-K. Ki, H. Chen, H. Berger, A. H. MacDonald, A. F. Morpurgo, Nat. Commun. \textbf{6}, 8339 (2015).
\bibitem{aref1}
W.-T. Lu,  Q.-F. Sun, and Q. Cheng,
Phys. Rev. B \textbf{105}, 125425 (2022).
\bibitem{Zubair}
M. Zubair, P. Vasilopoulos, and M. Tahir, Phys. Rev. B \textbf{101}, 165436 (2020).
\bibitem{Cysne}
T. P. Cysne, J. H. Garcia, A. R. Rocha, and T. G. Rappoport, Phys. Rev.B \textbf{97}, 085413 (2018).
\bibitem{QCheng2}
Q. Cheng, and Q.-F. Sun, Phys. Rev. B \textbf{105}, 165427 (2022).
\bibitem{Asano}
Y. Asano, T. Yoshida, Y. Tanaka, and A. A. Golubov, Phys. Rev. B \textbf{78}, 014514 (2008).
\bibitem{Hou}
Z. Hou, and Q.-F. Sun, Phys. Rev. B \textbf{96}, 155305 (2017).
\bibitem{QCheng3}
Q. Cheng, Z. Hou, and Q.-F. Sun, Phys. Rev. B \textbf{101}, 094508 (2020).
\bibitem{BTK}
G. E. Blonder, M. Tinkham, and T. M. Klapwijk,
Phys. Rev. B \textbf{25}, 4515 (1982).
\bibitem{QCheng4}
Q. Cheng and Q.-F. Sun, Phys. Rev. B \textbf{103}, 144518 (2021).

\end{thebibliography}
\end{document}